\newcommand{\be}{\begin{equation}}
\newcommand{\ee}{\end{equation}}
\newcommand{\bea}{\begin{eqnarray}}
\newcommand{\eea}{\end{eqnarray}}
\newcommand{\nn}{\nonumber}

\input epsf.sty

\documentclass[12pt]{iopart}

\begin{document}

\title[Two dimensional nonlinear cylindrical  
 equilibria...]
{Two dimensional  nonlinear  cylindrical
 equilibria with reversed magnetic shear and sheared flow}

\author{Ap Kuiroukidis$^{1}$ and G. N. Throumoulopoulos$^{2}$}

\address{$^{1}$Technological Education Institute of Serres 621 24 Serres, Greece
\\
$^{2}$Department of Physics, University of Ioannina, Association Euratom-Hellenic Republic,
 451 10 Ioannina, Greece}

\ead{kouirouki@astro.auth.gr,$\; \; $gthroum@cc.uoi.gr }

\begin{abstract}
Nonlinear tranlational symmetric equilibria with  up to
quartic flux terms in the free functions, reversed magnetic
shear and sheared flow are constructed in two ways:
i) quasianalytically by an ansatz which reduces the pertinent
generalized Grad-Shafranov equation to a set of ordinary
differential equations and algebraic constraints which is
then solved numerically, and ii) completely numerically by
prescribing analytically a boundary having  an X-point. The
equilibrium characteristics are then  examined by means of the
pressure, safety factor, current density and  electric field.
For  flows  parallel to the magnetic field the stability of the
equilibria constructed is also examined by applying a sufficient
condition. It turns out that the equilibrium nonlinearity has
a stabilizing impact which is slightly enhanced by the sheared flow.
In addition, the results indicate that the stability is affected
by the up-down asymmetry.

\end{abstract}

\section{Introduction}

 Sheared flows play a role in the transitions to  improved confinement regimes
 in magnetic confinement devices, as the L-H transition and the formation of
 internal transport barriers (ITBs), though  understanding  the physics of these
 transitions  remains incomplete.  In particular magnetohydrodynamic equilibria
 with flow, which is the basis of stability and transport studies, have been constructed
 as solutions to
generalized Grad-Shafranov equations, e.g. Eq. (\ref{eqpsi}) below,   \cite{mape1}-\cite{kuth3}.
In connection with the present study we refer  to our recent contribution \cite{kuth3}
in which up-down symmetric nonlinear two dimensional cylindrical equilibria with
incompressible flow pertinent to the L-H transition were obtained. Equilibria
relevant to the L-H tranistion usually have peaked toroidal current density profiles
and   safety factors increasing monotonically from the magnetic axis to the plasma boundary.

A necessary requirement for tokamak operation in connection with the ITER and DEMO projects  is a constant
toroidal plasma current, which produces the poloidal component of
the magnetic field. Among the different options
for such non-inductive current drive  (e.g.  electron
cyclotron current drive,  neutral beam current drive, bootstrap
current) only  the bootstrap   current can
produce a sufficiently large amount of toroidal current in
big tokamaks. The amount of bootstrap current
is proportional to the pressure gradient. Typically the
maximal pressure gradients are located off-axis thus leading
to  hollow current profiles in the plasma associated with reversed magnetic shear.
 Static equlibria with reversed magnetic shear was the subject of \cite{be}-\cite{maro}.

The stability of  fluids and plasmas in the presence of equilibrium flows non parallel to the
magnetic field remains a tough problem reflecting to the lack of necessary and sufficient conditions. Only
for parallel  flows few sufficient conditions for linear stability are available  \cite{FrVi}-\cite{throu3}.
In previous studies
we found that the stability condition of  \cite{throu3} is not
satisfied for the linear equilibria of \cite{apost} and
{\cite{ThTa3}  while it is satisfied within  an appreciable part of
the plasma  for the nonlinear equilibria of \cite{ttp},
\cite{throu4} and \cite{kuth3}.   This led us to the conjecture that the equilibrium
nonlinearity may act synergetically with the  sheared flow to
stabilize the plasma.

Aim of the present study  is to extend our previous paper \cite{kuth3} in two respects: up-down asymmetry
and reversed magnetic shear. As in  \cite{kuth3}  non magnetic field aligned      equilibrium flows will
be included.  In this respect it is noted that a synergism of reversed magnetic shear and
 sheared  poloidal and toroidal rotation,
 consisting in that on the one hand the reversed magnetic shear
  plays a role in triggering the ITBs development while on the other  hand the sheared rotation has an impact on the
  subsequent growth and allows the formation of strong ITBs,  was observed in JET \cite{devr} and
  DIII-D \cite{shmc}. In addition here the above conjecture about a combination of stabilizing effects
  of equilibrium nonlinearity and plasma flow will be checked.  The reason for considering  translational
symmetry is the many free physical and geometrical parameters
involved in connection with the flow amplitude,  direction and
shear, equilibrium nonlinearity, symmetry and toroidicity. Thus, in
the presence of nonlinearity one first could exclude toroidicity.

The organization of the paper is as follows:
In the second  section we briefly  present the
general setting for the translatinally symmetric equilibrium equations with incompressible flow  and
introduce the ansatz reducing the problem to a set of ordinary differential equations (ODEs)  and algebraic
constraints. In section 3  up-down asymmetric  equilibria are constructed quasianalytically and their
characteristics are studied. In section 4 equilibria with a lower X-point are derived  numerically by
imposing analytically the boundary shape. A stability consideration of the equilibria obtained  is made
in section 5. Section 6 summarizes the conclusions. 

\section{Translational symmetric equilibria with flow}

The equilibrium of a cylindrical plasma with incompressible
flow and arbitrary cross-sectional shape satisfies the generalized
Grad-Shafranov equation \cite{throu1}, \cite{simi}
\bea
\label{eqpsi}
(1-M_{p}^{2})\nabla^{2}\psi-\frac{1}{2}(M_{p}^{2})^{'}
|\nabla\psi|^{2}+\frac{d}{d\psi}
\left(\mu_{0}P_{s}+\frac{B_{z}^{2}}{2}\right)=0
\eea
for the poloidal magnetic flux function $\psi.$ Here,
$M_{p}(\psi),\; P_{s}(\psi),\; \rho(\psi)$ and $B_{z}(\psi)$
are respectively the poloidal Alfv\'{e}n Mach function, pressure
in the absence of flow, density and magnetic field parallel to
the symmetry axis $z$, which are surface quantities. Because
of the symmetry, the equilibrium quantities are $z$-independent and
the axial velocity $v_{z}$ does not appear explicitly in
Eq. (\ref{eqpsi}).  SI units are employed unless otherwise stated (see section 5).
Derivation of Eq. (\ref{eqpsi}) is based on the
following two steps: First express the divergence free fields
in terms of scalar quantities as
\bea
\label{fields}
{\bf B}&=&B_{z}\nabla z+\nabla z\times \nabla \psi \nn \\
\mu_{0}{\bf j}&=&\nabla^{2}\psi\nabla z-\nabla z\times\nabla B_{z}\nn \\
\rho{\bf v}&=&\rho v_{z}\nabla z+\nabla z\times\nabla F
\eea
The  velocity $\bf v$ relates to the electric field,  ${\bf E}=-\nabla\Phi$ (where $\Phi (\psi)$
is the electrostatic potential), by  Ohm's law, 
${\bf E}+{\bf v}\times{\bf B}=0$.
Second, project the momentum equation,
$\rho({\bf v}\cdot\nabla){\bf v}={\bf j}\times{\bf B}-\nabla P$,
and Ohm's law, along the symmetry
direction $z$, ${\bf B}$ and $\nabla\psi$. The projections yield
four first integrals in the form of surface quantities
(two out of which are $F(\psi)$ and $\Phi(\psi)$), Eq. (\ref{eqpsi})
and the Bernoulli relation for the pressure
\bea
\label{piesn}
P=P_{s}(\psi)-\frac{1}{2\mu_{0}}M_{p}^{2}(\psi)|\nabla\psi|^{2}
\eea
Because of the flow $P$ is not a surface quantity. Also the density
becomes surface quantity because of incompressibility and
$M_{p}^{2}(\psi)=(F^{'}(\psi))^{2}/(\mu_{0}\rho)$. Five of the
surface quantities, chosen here to be $P_{s},\; \rho,\; B_{z}, M_{p}^{2}$
and $v_{z}$, remain arbitrary. 

Using the mapping
\bea
\label{trans}
u(\psi)=\int_{0}^{\psi}[1-M_{p}^{2}(g)]^{1/2}dg,\; \; \; (M_{p}^{2}<1)
\eea
Eq. (\ref{eqpsi}) is transformed to
\bea
\label{eqmot1}
\nabla^{2}u+\frac{d}{du}
\left(\mu_{0}P_{s}+\frac{B_{z}^{2}}{2}\right)=0
\eea
Note that transformation (\ref{trans}) does not affect the
magnetic surfaces, it just relabels them. Eq. (\ref{eqmot1}) is
identical in form with the static equilibrium equation.

In the
present study we assign the free function term in Eq. (\ref{eqmot1})
as
\bea
\label{free}
\left(\mu_{0}P_{s}+\frac{B_{z}^{2}}{2}\right)=c_{0}+c_{1}u+c_{2}\frac{u^{2}}{2}
+c_{3}\frac{u^{3}}{3}+c_{4}\frac{u^{4}}{4}
\eea
to obtain
\bea
\label{eqmot2}
u_{xx}+u_{yy}+c_{1}+c_{2}u+c_{3}u^{2}+c_{4}u^{3}=0
\eea
where $(x,y)$ are the usual cartesian coordinates. The form of this equation
leads us to introduce the following up-down asymmetric ansatz for the flux function which enables reduction of 
the equilibrium problem  to a set of ordinary differential equations and first-order constraints:

\bea
\label{ansatz}
u=\frac{N_{2}(x)y^{2}+N_{1}(x)y+f(x)D_{0}(x)}{y^{2}+D_{1}(x)y+D_{0}(x)}
\eea
This is an extension of the respective ansatz for up-down symmetric equilibria we introduced for the first time in \cite{kuth3}.
Inserting Eq. (\ref{ansatz})  into  (\ref{eqmot2}), after a rather lengthy  calculation the latter is transformed into a fraction (F), the nominator
of which is a polynomial of $y$ of sixth order. Equating this nominator  to zero,
from the $y^{6}$-term we obtain
\bea
\label{eqmotn2}
N_{2}^{''}+c_{1}+c_{2}N_{2}+c_{3}N_{2}^{2}+c_{4}N_{2}^{3}=0
\eea
From the $y^{0}$-term it follows
\bea
\label{eqmotf}
f^{''}+\frac{2(N_{2}-f)}{D_{0}}-\frac{2D_{1}(N_{1}-fD_{1})}{D_{0}^{2}}+
c_{1}+c_{2}f+c_{3}f^{2}+c_{4}f^{3}=0
\eea
From the $y,\; y^{4},\; y^{5}$-terms one yields
\bea
\label{eqmotn1}
N_{1}^{''}&=&\frac{1}{DD}
\left[-N_{2}(N_{2}-f)G_{1}+(fD_{0}^{2}(N_{2}-f)-\right.\nn \\
& &\left.-D_{0}(N_{1}-fD_{1})(N_{1}+N_{2}D_{1}))
G_{5}+D_{0}N_{2}(N_{1}-fD_{1})G_{4}\right]
\eea
\bea
\label{eqmotd1}
D_{1}^{''}&=&\frac{1}{DD}
\left[-(N_{2}-f)G_{1}+(D_{0}^{2}(N_{2}-f)-2D_{0}D_{1}(N_{1}-fD_{1}))G_{5}\right.\nn
\\
& &\left.+D_{0}(N_{1}-fD_{1})G_{4}\right]
\eea
\bea
\label{eqmotd0}
D_{0}^{''}&=&\frac{1}{DD}
\left[(N_{1}-N_{2}D_{1})G_{1}-D_{0}^{2}(N_{1}+N_{2}D_{1}-2fD_{1})G_{5}\right.\nn
\\
& &\left.+D_{0}^{2}(N_{2}-f)G_{4}\right]
\eea
where $DD=D_{0}(N_{1}-fD_{1})(N_{1}-N_{2}D_{1})+D_{0}^{2}(N_{2}-f)^{2}$ and the
functions $G_{1},\; G_{4},\; G_{5}$ are given in the Appendix.
From the $y^{3},\; y^{2}$-terms we obtain respectively the first-order
constraints $C_{1},\; C_{2}$
\bea
\label{c1}
& &(D_{1}^{2}+2D_{0})N_{1}^{''}(\phi,\phi^{'})-(N_{1}+N_{2}D_{1}-2fD_{1})D_{0}^{''}(\phi,\phi^{'})
-\nn \\&-&
(N_{2}D_{0}+N_{1}D_{1}+fD_{0})D_{1}^{''}(\phi,\phi^{'})+G_{3}=0
\eea
and
\bea
\label{c2}
& &2D_{0}D_{1}N_{1}^{''}(\phi,\phi^{'})-(D_{0}(N_{2}-f)+D_{1}(N_{1}-fD_{1}))D_{0}^{''}(\phi,\phi^{'})
-\nn \\&-&D_{0}(N_{1}+fD_{1})D_{1}^{''}(\phi,\phi^{'})+G_{2}=0
\eea
Here $\phi$ denotes collectively all the functions appearing in Eq. (\ref{ansatz}), and
Eqs. (\ref{eqmotn1}-\ref{eqmotd0}) are used. The functions $G_{2},\; G_{3}$ are also given
in the Appendix. Note that Eqs. (\ref{c1}), (\ref{c2}) are not second order ``evolution" differential  equations but
rather first-order constraints to be fulfilled during numerical integration. This justifies   the introduction of  ansatz (\ref{ansatz}) which results in a simple numerical treatment of the equilibrium  through ordinary differential equations and algebraic constraints 
in contrast and alternative to the full numerical treatment of section 4.


\section{Class of quasianalytic  solutions}

The system of Eqs. (\ref{eqmotn2}-\ref{eqmotd0}) is integrated numerically
using high-precision numerical integration with very small step size, due to
its extreme complexity and nonlinearity.
The magnetic axis regularized with respect to the geometric center is taken to
be ``Shafranov shifted"\footnote{The term  ``Shafranov shift"  here  means  that
the equilibrium in addition to up-down is left-right asymmetric  and is not connected   to the
toroidicity which vanishes in cylindrical geometry.} at $x_{a}=1+x_{s}=1.1$.
We take the following ITER-pertinent geometrical data:  $a=2\ m$, $R_{0}=6.2\ m$
for the minor and major radius of
the ``torus" respectively and the inverse aspect ratio is
$\epsilon _{0}=0.32$. The bounds for the $x-$variable are
$x_{min}=1-\epsilon_{0}$ and $x_{max}=1+\epsilon_{0}$. The integration
begins from $x_{a}$ forward up to $x_{max}$ and backwards up to $x_{min}$.
For the following values of the parameters $c_{1}=52.0$, $c_{2}=-0.4$, $c_{3}=0.1$
and $c_{4}=0.1$ and for initial conditions $N_{2}=-1.4$, $N_{2}^{'}=0.95$,
$N_{1}=-0.15$, $N_{1}^{'}=-1.05$, $D_{1}=0.15$, $D_{1}^{'}=-0.137$,
$D_{0}=1.05$, $D_{0}^{'}=0.1$, $f=7.5$ and $f^{'}=0.0$ we obtained the
solution of Fig. 1. The first-order constraints of Eqs. (\ref{c1}-\ref{c2}) were
monitored during the  integration and they were kept to very low values. The
product of the constraint $C_{1}$ with the average value of $y^{3}$ (taken to
be equal to 0.2) and the constraint $C_{2}$ with the average value of $y^{2}$,
that appear in the nominator of the final fraction (F)
(see the text just below Eq. (\ref{ansatz}))
were kept bounded to  $|C_{1}|\leq 0.07$ and $|C_{2}|\leq 0.18$. This combined
with the fact that the denominator of this fraction has a positive definite value
greater or equal to 1.0, is an additional argument that this fraction is very close
to zero and that the presented equilibrium is indeed an acceptable solution of
Eq. (\ref{eqmot2}).

The bounding surface (shown in green) corresponds to $u_{b}=5.5\ Wb$ while the
magnetic axis  to the value $u_{a}=7.8\ Wb$. The magnetic axis
is located at $(x_{a},y_{a})=(1.1,-0.07)$. A quartic fitting yields expressions
for the functions $N_{2}, N_{1}, D_{1}, D_{0}, f$ such as the following equation
\bea
\label{fit}
N_{2}&=&5.527x^{4}-23.78x^{3}+11.81x^{2}+31.95x-27.27
\eea
We stress however the fact that for a correct representation and plotting
of the various functions all the higher order  expansions and more precise
numerical parametric values  are needed. Plotting Eq. (\ref{ansatz}) using
Eq. (\ref{fit}) will {\it not} yield the correct result of Fig. 1, which
occurs through the precise numerical results for these functions.

The  MHD safety factor is defined as
 \cite{wess}
\bea
\label{sfa}
q(u)=\frac{d\psi _{tor}/dV}{d\psi _{pol}/dV}=
\frac{1}{2\pi }\frac{d\psi _{tor}/dV}{d\psi /dV}
\eea
Using $dV=2\pi |J|d\psi d\theta $,
$J^{-1}=\nabla \theta \cdot (\nabla \phi \times \nabla \psi )$ and
expressing the
fluxes in terms of the magnetic field, one can cast
(\ref{sfa}) in the form of a line integral on each constant$-u$ curve. The
detailed evaluation of $q$ for the present equilibrium  of Fig. 1
yields the curve of Fig. 2 with strong reversed magnetic shear. For this result it was used $c_{0}=200.0$
in Eq. (\ref{free}). Also for the axial magnetic field it was adopted the
typical tokamak diamagnetic function $B_{z}=B_{z0}(1+\gamma(1-\frac{u}{u_{b}}))$,
shown in Fig. 3, with $\gamma=0.1$
and $B_{z0}=3.2\ T$ . Then, the static pressure function   $P_{s}(u)$ is computed by Eq. (\ref{free}),
while for the flow
function in Eq. (\ref{trans}) it was used
$M_{p}^{2}=M_{pa}(\frac{u}{u_{b}}-1)^{2.5}$ with $M_{pa}=0.1$
and $u_{b}=5.5\ Wb \leq u\leq u_{a}=7.5\ Wb$. The pressure (Eq. (\ref{piesn}))
is shown in Fig. 4, normalized to its center value of $P_{0}=2.1046\times 10^{5\ }Pa$.  Also
instead of the axial velocity $v_{z}$, the corresponding Mach function $M_{z}^{2}$
is chosen similar to the poloidal one ($M_{z}^{2}\simeq M_{p}^{2}$) with $M_{za}=1.1M_{pa}$.

The electric field for equilibrium of Fig. 1 is shown in Fig. 5.
Here the choice $\rho=\rho_{a}(\frac{u}{u_{b}}-1)^{0.5}$ has
been made for the density  with $\rho_{a}=4.0\times 10^{-7}Kgr\; m^{-3}$.
The maximum of $\bf E$ increases with the flow parameter $M_{pa}$ but
the position of the maximum is not affected by the flow in agreement with the
results of \cite{simi,kuith2}. The hollow axial current density profile in the midplane
$y=0$ is shown in Fig. 6 in consistence  with the negative
magnetic shear curve for the safety factor of Fig. 2.


\begin{figure}[ht!]
\centerline{\mbox {\epsfxsize=10.cm \epsfysize=8.cm \epsfbox{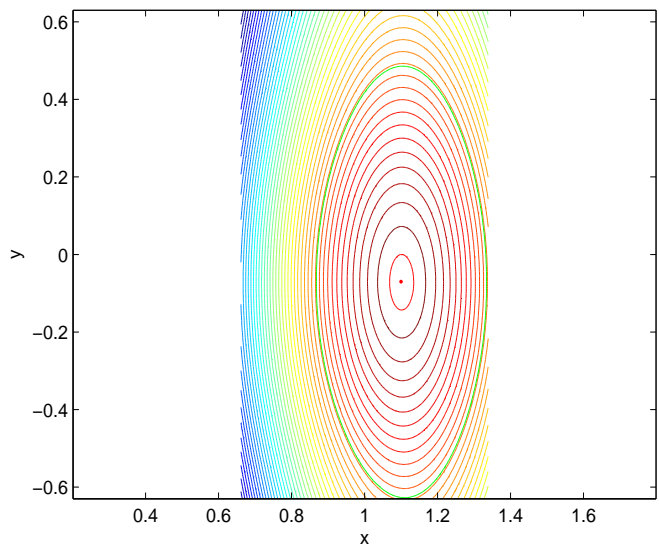}}}
\caption[]{Equilibrium for the initial conditions given  in the text of section 2. The
bounding surface, shown in green, corresponds to $u_{b}=5.5\ Wb$ while at the
magnetic axis, which is located at $(x_{a},y_{a})=(1.1,-0.07)$, $u_{a}=7.8\ Wb$.}
\label{fig1}
\end{figure}

\begin{figure}[ht!]
\centerline{\mbox {\epsfxsize=10.cm \epsfysize=8.cm \epsfbox{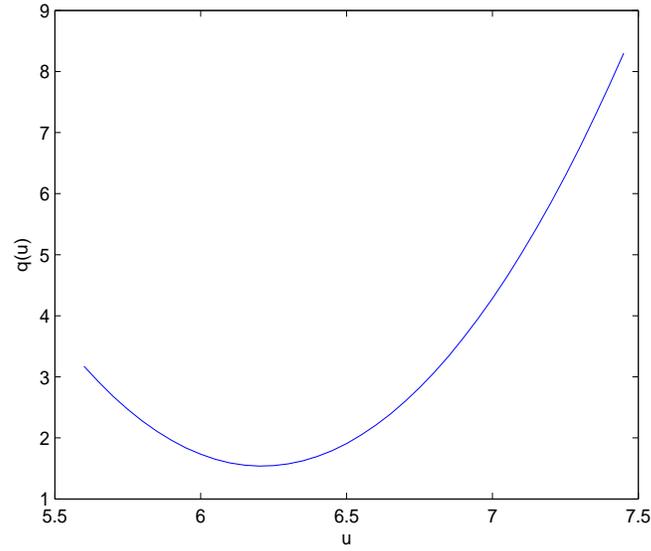}}}
\caption[]{The safety factor for the equilibrium of Fig. 1 presenting a strong negative magnetic shear region.  The outer
bounding surface corresponds to $u=u_{b}=5.5\ Wb$ while at the magnetic axis $u=u_{a}=7.5\ Wb$}
\label{fig2}
\end{figure}

\begin{figure}[ht!]
\centerline{\mbox {\epsfxsize=10.cm \epsfysize=8.cm \epsfbox{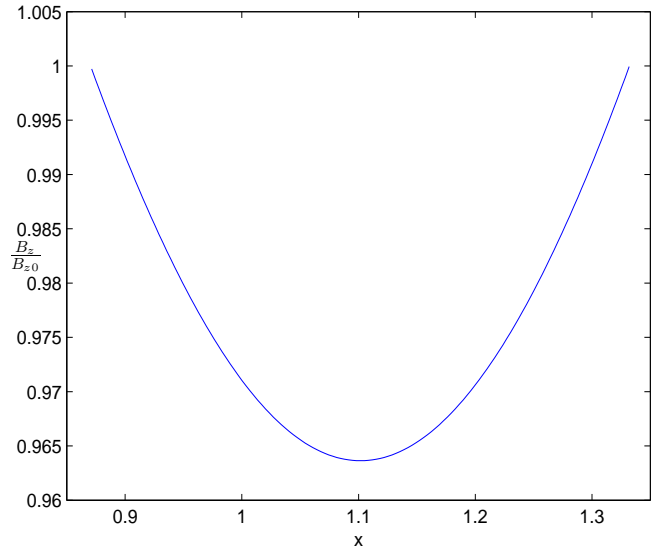}}}
\caption[]{The axial magnetic field $B_{z}$, for equilibrium of Fig. 1
normalized to its center value of $B_{z0}=3.2\ T$.}
\label{fig3}
\end{figure}

\begin{figure}[ht!]
\centerline{\mbox {\epsfxsize=10.cm \epsfysize=8.cm \epsfbox{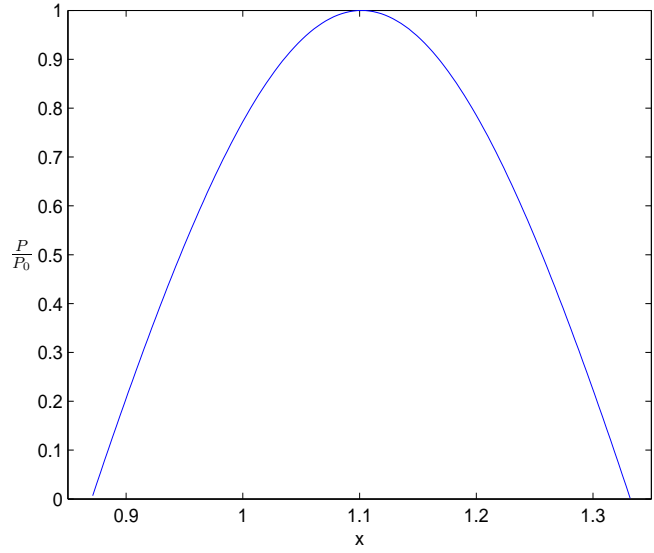}}}
\caption[]{The pressure for equilibrium of Fig. 1 normalized to its center value
of $P_{0}=2.1046\times 10^{5}Pa$.}
\label{fig4}
\end{figure}

\begin{figure}[ht!]
\centerline{\mbox {\epsfxsize=10.cm \epsfysize=8.cm \epsfbox{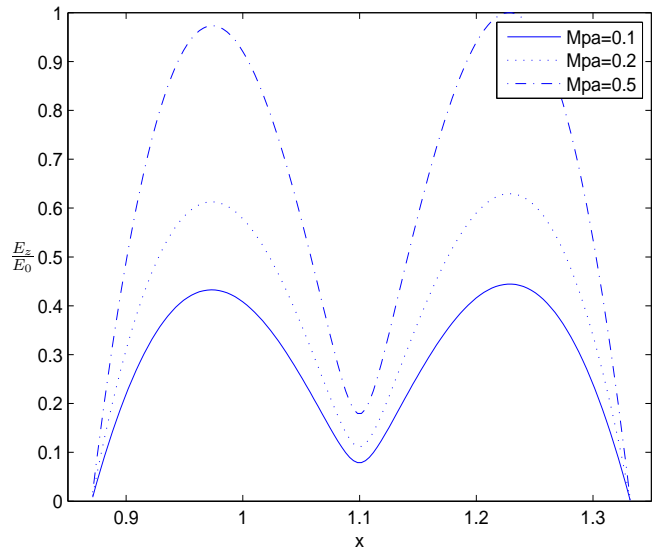}}}
\caption[]{The  electric field $E_{z}$, for equilibrium of Fig. 1
normalized to its maximum value of $E_{0}=5.4851\; kV\; m^{-1}$.}
\label{fig5}
\end{figure}

\begin{figure}[ht!]
\centerline{\mbox {\epsfxsize=10.cm \epsfysize=8.cm \epsfbox{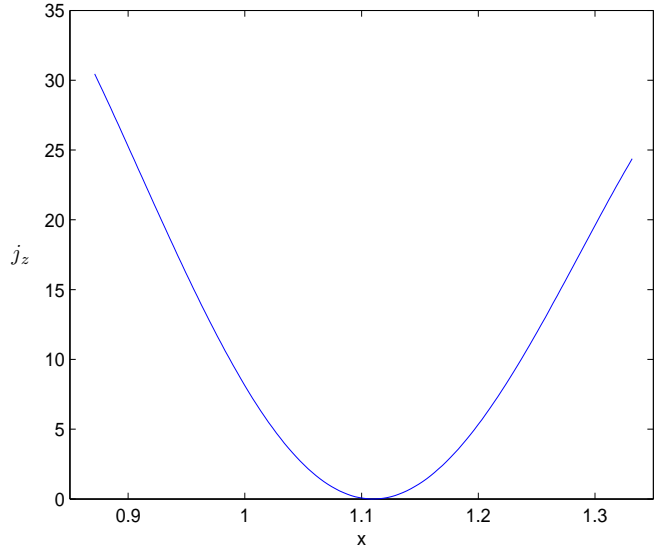}}}
\caption[]{The axial current density at the midpalne $A(x,y=0)$ for the
equilibrium of Fig. 1. It is hollow in connection with the negative
magnetic shear of Fig. 2.}
\label{fig6}
\end{figure}

\begin{figure}[ht!]
\centerline{\mbox {\epsfxsize=10.cm \epsfysize=8.cm \epsfbox{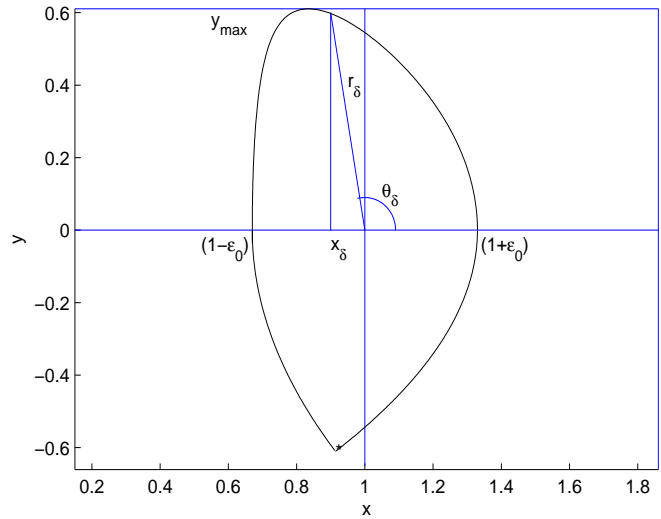}}}
\caption[]{Bounding flux surface for the asymmetric case defined by Eqs. (\ref{upbound}-\ref{right}),
possesing a divertor, null X-point at $P_{X}(x_{X},y_{X})=(0.9139,-0.6105)$.}
\label{fig7}
\end{figure}

\begin{figure}[ht!]
\centerline{\mbox {\epsfxsize=10.cm \epsfysize=8.cm \epsfbox{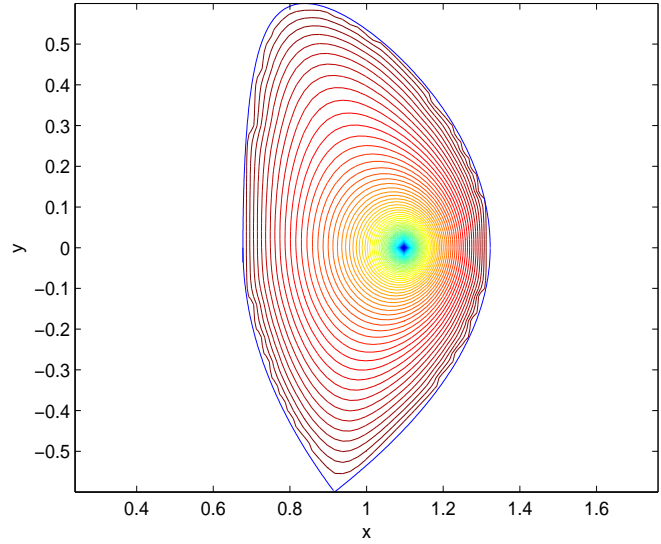}}}
\caption[]{The numerical solution of Eq. (\ref{eqmot2}) for the parameter
values given in the text of section 4.}
\label{fig8}
\end{figure}

\begin{figure}[ht!]
\centerline{\mbox {\epsfxsize=10.cm \epsfysize=8.cm \epsfbox{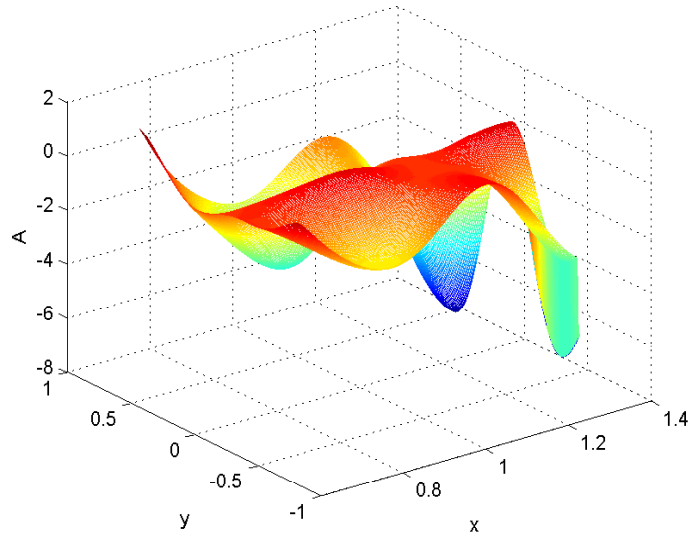}}}
\caption[]{The stability function $A$ for the equilibrium of Fig. 1. At the upper
part $(y>0)$ of the equilibrium it mostly assumes positive values, while at the
lower part $(y<0)$ it assumes negative values.}
\label{fig9}
\end{figure}

\begin{figure}[ht!]
\centerline{\mbox {\epsfxsize=10.cm \epsfysize=8.cm \epsfbox{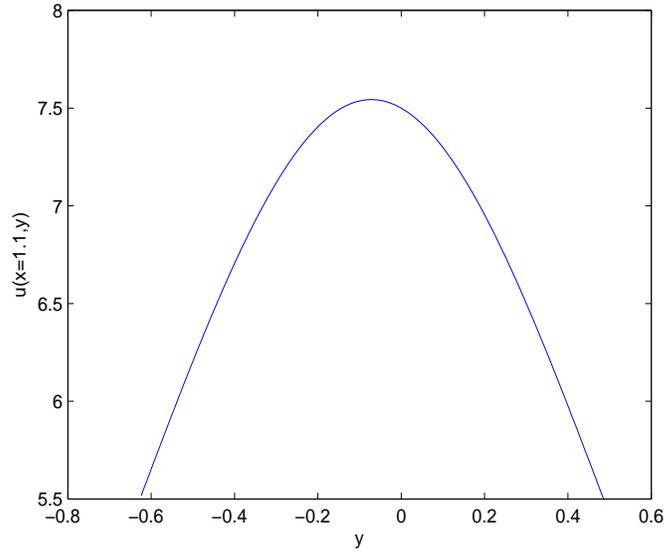}}}
\caption[]{The poloidal magnetic flux function $u(x=1.1,y)$ for the
equilibrium of Fig. 1. It is slightly up-down asymmetric with repsect to
$y=0$.}
\label{fig10}
\end{figure}

\begin{figure}[ht!]
\centerline{\mbox {\epsfxsize=10.cm \epsfysize=8.cm \epsfbox{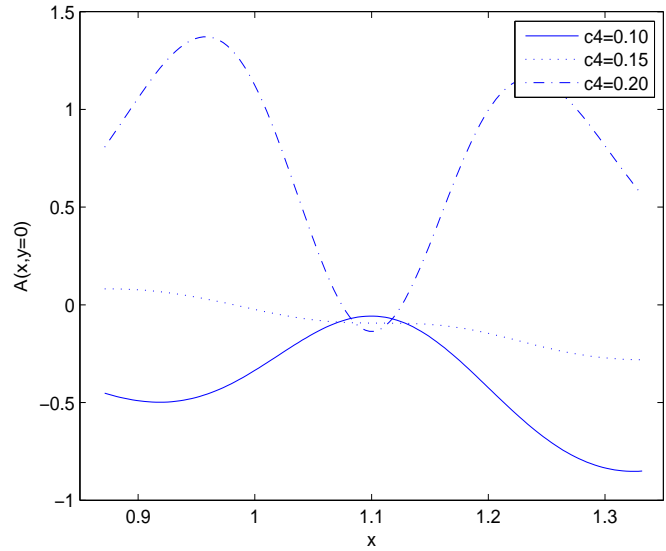}}}
\caption[]{The stability function at the midpalne $A(x,y=0)$ for the
equilibrium of Fig. 1, for various values of the nonlinear constant $c_{4}$.
It appears that nonlinearity acts in favour of the stability.}
\label{fig11}
\end{figure}

\begin{figure}[ht!]
\centerline{\mbox {\epsfxsize=10.cm \epsfysize=8.cm \epsfbox{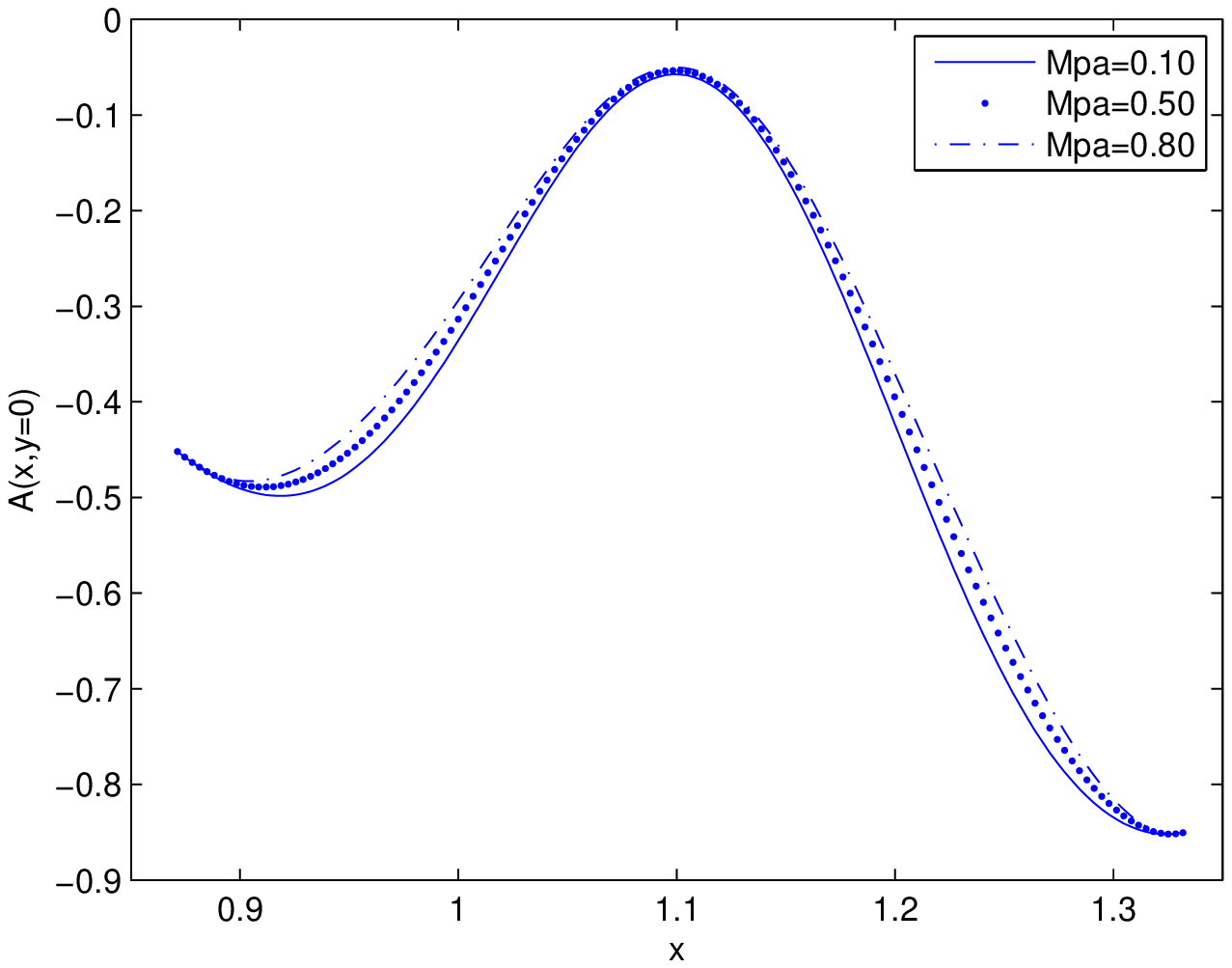}}}
\caption[]{The stability function at the midplane $A(x,y=0)$ for the
equilibrium of Fig. 1, for various values of the flow parameter $M_{pa}$.
It appears that flow acts in favour of the stability, though in
a weak manner.}
\label{fig12}
\end{figure}



\section{Numerical asymmetric equilibrium with X-point}

We consider now the direct numerical solution of Eq. (\ref{eqmot2})
with a prescribed boundary possessing a divertor null X-point.
We first will specify  the boundary. The boundary conditions for the
flux function $u$ is $u_{b}=1\ Wb$ on the prescribed boundary
curve and $u_{a}=0\ Wb$ on the magnetic axis. The magnetic axis is
taken to be Shafranov shifted at $x_{a}=1+x_{s}=1.1$. The model thus
has three free parameters which are the Shafranov shift $x_{s}$,
the elongation $\kappa$ and the triangularity $\delta.$ Their
values are taken as $x_{s}=0.1$, $\kappa=1.86$ and $\delta=0.5$,
in accordance with the corresponding data of the ITER project. The
bounding flux surface, for the aymmetric case, is shown in Fig. 7.
We take the values  $a=2\ m$, $R_{0}=6.2\ m$ for the minor and major radius of
the ``torus" respectively for which  the inverse aspect ratio is
$\epsilon _{0}=0.32$.

The equation for the upper part of the bounding flux surface,
which if taken to hold for the lower part as well would give a symmetric
bounding surface, is
\bea
\label{upbound}
x_{b}&=&1+\epsilon _{0}cos(\tau +\alpha sin(\tau))\nn \\
y_{b}&=&y_{max}sin(\tau)
\eea
where $y_{max}=\kappa\epsilon _{0}$ with $\delta=(1-x_{\delta})/\epsilon_{0}$,
and $\alpha=sin^{-1}(\delta)$. Thus the following relations hold:
$x_{\delta}=1-\delta\epsilon_{0}$ and $\theta_{\delta}=\pi-tan^{-1}(\kappa/\delta)$.
The parameter $\tau$ is any increasing function of the polar angle
$\theta$, satisfying $\tau(0)=0$, $\tau(\pi)=\pi$ and $\tau(\theta_{\delta})=\pi/2$.
In our model we take
\bea
\label{taf}
\tau(\theta)&=&t_{0}\theta^{2}+t_{1}\theta^{n}\nn \\
t_{0}&=&\frac{\theta_{\delta}^{n}-\frac{1}{2}\pi^{n}}{\pi\theta_{\delta}^{n}-\theta_{\delta}^{2}\pi^{n-1}}
\nn \\
t_{1}&=&\frac{-\theta_{\delta}^{2}+\frac{1}{2}\pi^{2}}{\pi\theta_{\delta}^{n}-\theta_{\delta}^{2}\pi^{n-1}}
\eea
with $n=8$. In order to complete the asymmetric bounding curve we specify now
the lower part of it $(y<0)$ as follows. The left lower  branch of the curve is given by
\bea
\label{left}
x_{b}&=&1+\epsilon_{0}cos(\theta)\nn \\
y_{b}&=&-[2p_{1}\epsilon_{0}(1+cos\theta)]^{1/2}\nn \\
p_{1}&=&\frac{y_{max}^{2}}{2\epsilon_{0}(1+cos\theta_{\delta})},\; \; \; \;
(\pi\leq \theta \leq 2\pi-\theta_{\delta})
\eea
while the right  lower branch of the curve is given by
\bea
\label{right}
x_{b}&=&1+\epsilon_{0}cos(\theta)\nn \\
y_{b}&=&-[2p_{2}\epsilon_{0}(1-cos\theta)]^{1/2}\nn \\
p_{2}&=&\frac{y_{max}^{2}}{2\epsilon_{0}(1-cos\theta_{\delta})},\; \; \; \;
(2\pi-\theta_{\delta}\leq \theta \leq 2\pi)
\eea
The divertor null X-point is located at $x_{X}=1+\epsilon_{0}cos\theta_{\delta}=0.9139$
and $y_{X}=-y_{max}=-0.6105$.

The Laplacian operator $u_{xx}+u_{yy}$ is discretized on a
rectangular grid where we have\\ $(1-\epsilon_{0})\leq x\leq (1+\epsilon_{0})$
and $-y_{max}\leq y\leq y_{max}$ with grid step $h$. The nine point formula
is then employed \cite{numer}
\bea
\label{ninepoint}
\nabla^{2}u_{i,j}&=&\frac{1}{6h^{2}}
\left[u_{i+1,j+1}+4u_{i+1,j}+u_{i+1,j-1}+4u_{i,j+1}+4u_{i,j-1}+u_{i-1,j+1}+\right.\nn \\
& &\left.+4u_{i-1,j}+u_{i-1,j-1}-20u_{i,j}\right]
\eea
and when substituted into Eq. (\ref{eqmot2}), the latter is written as
$u_{i,j}^{(new)}=G(u)_{i,j}^{(old)}$ and is solved iteratively. The last
term of Eq. (\ref{ninepoint}) is taken to be the $u_{i,j}^{(new)}$ and
the iterations stop (i.e. reaching  convergence to the solution) when
$|u_{i,j}^{(new)}-u_{i,j}^{(old)}|<0.001$. We stress again that the
 conditions $u_{b}=1$ on the boundary and $u_{{a}}=0$ on axis are imposed in
every iteration. For $h=0.02$, and for the following values of the
constants of Eq. (\ref{eqmot2}), $c_{1}=-10.0,\; c_{2}=2.0,\; c_{3}=1.1,
c_{4}=1.1$, a number of $N=165$ iterations were needed to obtain the
desired accuracy. The solution is shown in Fig. 8. Although it seems
that the solution is dependent on the specific value of the $h$
chosen, from the discrete two-dimensional matrix of $u_{i,j}$ produced,
a two-dimensional Lagrange fitting is performed that yields a
polynomial in $(x,y)$. This is h-independent. It is noted that the ripples of the flux
function, appearing near the boundary of the equilibrium are due to
numerical instabilities that are inevitable present in the calculation.

\section{A stability consideration}
We now address  the important issue of the stability of the
solutions constructed with respect to small linear MHD
perturbations by means of the  sufficient condition of
\cite{throu3}. This condition concerning internal modes states that a general steady state of
a plasma of constant density and incompressible flow parallel to
$\bf B$ is linearly stable to small three-dimensional perturbations
if the flow is sub-Alfv\'enic ($M_p^2<1$) and $A\geq 0$, where $A$ is
given below by (\ref{cond}). Consequently, using henceforth
dimensionless quantities we set $\rho=1$. Also, for parallel flows
(${\bf v}=M{\bf B}$) it holds $M_p\equiv M_z\equiv M $. In fact if
the density is uniform at equilibrium it remains so at the perturbed
state because of incompressibility.
In the $u$-space for axisymmetric equilibria $A$ assumes the form
\bea
\label{cond} A&=&-{\bar g}^{2} \left[\frac{}{}({\bf j}\times \nabla u)\cdot ({\bf
B}\cdot\nabla)\nabla u+\right.
\nn \\
&+&\left(\frac{M_{p}^{2}}{2}\right)^{'}\frac{|\nabla u|^{2}}{(1-M_{p}^{2})^{3/2}}
\left\{\frac{}{}\nabla u\cdot \nabla (B^{2}/2)+\right. \nn \\
&+&\left.\left.{\bar g}\frac{|\nabla
u|^{2}}{(1-M_{p}^{2})^{1/2}}\frac{}{}\right\}\right]
\eea
with
$$\bar{g}:=\frac{P_{s}^{'}(u)-(M_{p}^{2})^{'}B^{2}/2}{1-M_{p}^{2}}$$
This condition, although complicated is accurate (a proof is provided in \cite{throu3}) and all the computations and
conclusions have been performed with great care.
Specifically, its application to  the equilibria constructed in sections 3 and 4 led to the following results:
\begin{enumerate}
\item Even a weak up-down asymmetry affects stability  as indicated in Fig. 9 where
 we have checked thoroughly the values of
the function $A$ for the equilibrium of Fig. 1.
It turns out that at most of the upper part of
the equilibrium, where $y>0$,  $A$ assumes small positive values, while for the
lower part of the equilibrium, $(y<0)$, it assumes small negative values.
This slight $A$-up-down asymmetry is connected to  the respective slight asymmetry of the flux function $u$
with respect to the vertical position $y$; the latter can be seen in Fig. 10 where the profile $u(x,y)$ is given at
the point $x=1+x_{s}=1.1$. Though $A<0$ does not necessarily imply an unstable equilibrium because the
condition is sufficient, the above result is consistent with the fact that up-down asymmetry may make the plasma unstable. For external modes this might  relate to 
the vertical instability, e.g.  \cite{fi}.
\item The non linearity  favours the stability  as it is shown in Fig. 11 where $A$ is plotted as
a function of $x$ at the midplane $y=0$ for  increasing values of the non-linearity
constant $c_{4}$.
\item The  flow has a slight stabilizing effect. An example is given in Fig. 12 in connection
with the flow parameter  $M_{pa}$.
\end{enumerate}

\section{Summary}
We have constructed and studied nonlinear translational symmetric equilibria with
strong  reversed magnetic shear and sheared incompressible flow non parallel to the
magnetic field on the basis of a generalized Grad-Shafranov equation
(Eq. (\ref{eqpsi})). This equation can be transformed to one identical in form  with the
static Grad-Shafranov equation which we have solved in a couple of  alternative ways:
 i) by using  an ansatz for the unknown magnetic flux function  (Eq. (\ref{ansatz}) which
 reduces the original equation to a set of ODEs and algebraic constraints
 (Eqs. (\ref{eqmotf}-\ref{c2})); then this set of equations is solved numerically, and ii)
 fully  numerically by prescribing anallyticaly a diverted boundary (Eqs. (\ref{upbound}-\ref{right})).
 The equilibria constructed are typically dimagnetic (Fig. 3), have peacked pressure
 profiles (Fig. 4), hollow toroidal current densities (Fig. 6) and electric fields
 possessing  a maximum (Fig. 5). The maximum of ${\bf E}$ takes larger values as the
 flow amplitude increases but its position is insensitive to the flow.

For parallel flows application of a condition for linear stability implies that
the equilibrium nonlinearity has a stabilizing effect together with a weaker
stabilizing impact of the sheared flow in agreement with past nonlinear equilibrium
studies \cite{ttp,throu4,kuth3}. Also even a small up-down asymmetry influences
stability.

 Finally it would be interesting to try constructing  equilibria with flow
 and reversed current density in connection with non nested magnetic surfaces
 thus generalizing  the static ones of \cite{mame,wa,robi} and extend the study
 to axially symmetric equilibria by possibly generalizing  the ansatz  (\ref{ansatz})
 in order to examining the impact of toroidicity.

\section*{Appendix: Quantities appearing in the ODEs and algebraic constraints (\ref{eqmotn1}-\ref{c2})}
The functions $G_{1},\; G_{4},\; G_{5}$ appearing into Eqs. (\ref{eqmotn1}),
(\ref{eqmotd1}), (\ref{eqmotd0}) are given by
\bea
\label{g1}
G_{1}&=&-4D_{0}D_{1}(N_{2}-f)+4D_{1}^{2}(N_{1}-fD_{1})-6D_{0}(N_{1}-fD-{1})
\nn \\
&-&2D_{0}D_{0}^{'}(N_{1}^{'}-fD_{1}^{'}-f^{'}D_{1})+2(D_{0}^{'})^{2}(N_{1}-fD_{1})
-2D_{0}^{2}D_{1}^{'}f^{'}+\nn \\&+&c_{1}D_{0}^{2}D_{1}+c_{2}D_{0}^{2}N_{1}
-c_{3}f^{2}D_{0}^{2}D_{1}+2c_{3}fD_{0}^{2}N1+\nn \\&+&3c_{4}f^{2}D_{0}^{2}N_{1}
-2c_{4}f^{3}D_{0}^{2}D_{1}
\eea
\bea
\label{g4}
G_{4}&=&-2(N_{2}-f)+\frac{2D_{1}(N_{1}-fD_{1})}{D_{0}}-2D_{0}^{'}(N_{2}^{'}-f^{'})
-\nn \\&-&2D_{1}^{'}(N_{1}^{'}+D_{1}N_{2}^{'}-N_{2}D_{1}^{'})+2c_{1}D_{1}^{2}
+2c_{2}N_{1}D_{1}+c_{3}N_{1}^{2}+\nn \\&+&2c_{3}fD_{0}N_{2}+2c_{3}N_{1}N_{2}D_{1}
-c_{3}f^{2}D_{0}-c_{3}N_{2}^{2}D_{1}^{2}-c_{3}D_{0}N_{2}^{2}-\nn
\\&-&c_{4}f^{3}D_{0}-c_{4}N_{2}^{3}D_{1}^{2}+3c_{4}fD_{0}N_{2}^{2}+
3c_{4}N_{2}N_{1}^{2}-2c_{4}N_{2}^{3}D_{0}
\eea
\bea
\label{g5}
G_{5}&=&-2D_{1}^{'}N_{2}^{'}+c_{1}D_{1}+c_{2}N_{1}+2c_{3}N_{1}N_{2}
-c_{3}N_{2}^{2}D_{1}-2c_{4}N_{2}^{3}D_{1}+\nn \\&+&3c_{4}N_{1}N_{2}^{2}
\eea
The functions $G_{2},\; G_{3}$ appearing into Eqs. (\ref{c1}-\ref{c2}) are
given by
\bea
\label{g2}
G_{2}&=&-2D_{1}^{2}(N_{2}-f)+\frac{2D_{1}^{3}(N_{1}-fD_{1})}{D_{0}}
+2c_{1}D_{0}D_{1}^{2}-c_{3}f^{2}D_{0}D_{1}^{2}-\nn \\&-&c_{4}f^{3}D_{0}D_{1}^{2}
-10D_{0}(N_{2}-f)+4D_{1}(N_{1}-fD_{1})-c_{3}f^{2}D_{0}^{2}-\nn \\&-&
2c_{4}f^{3}D_{0}^{2}+2f^{'}D_{0}^{'}(D_{1}^{2}+2D_{0})-2D_{0}^{'}(f^{'}D_{0}+fD_{0}^{'})
\nn \\&-&
2D_{1}D_{1}^{'}(f^{'}D_{0}+fD_{0}^{'})-c_{3}N_{2}^{2}D_{0}^{2}-c_{4}N_{2}^{3}D_{0}^{2}
-2N_{1}^{'}D_{0}^{'}D_{1}-\nn \\&-&2N_{1}^{'}D_{0}D_{1}^{'}-2N_{2}^{'}D_{0}D_{0}^{'}
+2fD_{0}(D_{1}^{'})^{2}+2N_{2}(D_{0}^{'})^{2}+4N_{1}D_{0}^{'}D_{1}^{'}+\nn
\\&+&2c_{2}N_{1}D_{0}D_{1}+2c_{3}fD_{0}D_{1}N_{1}+c_{3}N_{1}^{2}D_{0}+
2c_{3}fD_{0}^{2}N_{2}+3c_{4}fD_{0}N_{1}^{2}+\nn \\&+&3c_{4}f^{2}D_{0}^{2}N_{2}
\eea
and
\bea
\label{g3}
G_{3}&=&-4D_{1}(N_{2}-f)+\frac{4D_{1}^{2}(N_{1}-fD_{1})}{D_{0}}-2c_{3}f^{2}D_{0}D_{1}
-2c_{4}f^{3}D_{0}D_{1}+\nn \\&+&4D_{1}f^{'}D_{0}^{'}-2c_{3}N_{2}^{2}D_{0}D_{1}
-2c_{4}N_{2}^{3}D_{0}D_{1}-2(f^{'}D_{0}+fD_{0}^{'})D_{1}^{'}-\nn
\\&-&2N_{1}^{'}D_{0}^{'}-2N_{1}^{'}D_{1}D_{1}^{'}-2N_{2}^{'}D_{0}^{'}D_{1}-
2N_{2}^{'}D_{0}D_{1}^{'}+2N_{1}(D_{1}^{'})^{2}+\nn \\&+&4N_{2}D_{0}^{'}D_{1}^{'}+
2N_{1}-2N_{2}D_{1}+c_{1}D_{1}^{3}+2c_{1}D_{0}D_{1}+c_{2}N_{1}D_{1}^{2}+\nn
\\&+&2c_{2}N_{1}D_{0}+2c_{3}fD_{0}N_{1}+c_{3}N_{1}^{2}D_{1}+2c_{3}fD_{0}D_{1}N_{2}
+\nn \\&+&+2c_{3}N_{1}N_{2}D_{0}+c_{4}N_{1}^{3}+6c_{4}N_{1}N_{2}fD_{0}
\eea

\section*{Aknowledgments}\

One of the authors (GNT) would like to thank Henri Tasso and George Poulipoulis for useful discussions.

The work leading to this article was performed within
the participation of the University of Ioannina in the Association
Euratom-Hellenic Republic, which is supported in
part by the European Union (Contract of Association No.
ERB 5005 CT 99 0100) and by the General Secretariat of
Research and Technology of Greece. The views and opinions
expressed herein do not necessarily reflect those of the European
Commission.


\end{document}